\documentclass[adp,a4paper,fleqn%
]{w-art}
\usepackage{times,cite,w-thm}
\usepackage{amsmath,amssymb}
\theoremstyle{plain}

\theoremstyle{definition}

\usepackage[]{graphicx}
\begin{document}
\pagespan{1}{}
\Receiveddate{XXXX}
\Reviseddate{XXXX}
\Accepteddate{XXXX}
\Dateposted{XXXX}



\title[Thinking locally - reflections on DMFT.]{Thinking locally: reflections on Dynamical Mean-Field Theory from a high-temperature/high energy perspective.}


\author[A.~Georges]{Antoine Georges\inst{1,2}
\footnote{Corresponding author\quad E-mail:~\textsf{antoine.georges@polytechnique.edu},
           Phone: +33\,1\,69 33 40 21,
          Fax:+33\,1\,69 33 49 49}}
\address[\inst{1}]{Coll\`ege de France, 11 place Marcelin Berthelot, 75005 Paris}
\address[\inst{2}]{Centre de Physique Th\'eorique, Ecole Polytechnique, CNRS, 91128 Palaiseau Cedex, France}

\begin{abstract}
When spatial correlations are short-range, the physics of strongly correlated systems is controlled 
by local quantum fluctuations.  
In those regimes, Dynamical Mean-Field Theory can be viewed as a `compass' which provides 
guidance on the relevant degrees of freedom and their effective dynamics over intermediate energy scales. 
These intermediate energy scales and associated crossovers play a crucial role in the physics of strongly correlated materials. 
\end{abstract}
\maketitle                   





\section{Introduction}

Over the past two decades, Dynamical Mean-Field Theory (DMFT) has had considerable impact on 
our understanding of the physics of strongly correlated quantum systems, both in the context of 
simplified models and of real materials~\cite{ georges_review_dmft,kotliar_dmft_physicstoday} 

At the heart of this approach is the representation of the local spectral function of the solid as 
that of an atom self-consistently embedded into an effective medium.
This is a deep change of paradigm as compared to more traditional approaches in solid-state theory, 
which usually describe a solid as an inhomogeneous electron gas, to which interactions are later added. 
In contrast, DMFT gives central importance to the fact that, after all, solids are made of atoms (!) and that 
an isolated atom is a small many-body problem in itself which must be properly described from the start. 
In strongly-correlated materials, electrons are hesitant entities with a dual character: at high-energy they 
behave as localized, while at low-energy (in metallic systems) they may eventually form itinerant quasiparticles, 
albeit with a strongly suppressed spectral weight. In such systems, the local atomic point of view is therefore a 
particularly meaningful starting point, with a built-in real-space description of the atomic multiplets 
and of their broadening in the solid. As to low-energy quasiparticles, their description is made possible in 
this approach by the hybridization with the effective medium, which may screen the atomic multiplets 
through a Kondo-like process. 
The dual nature of the electron in strongly-correlated materials is also a manifestation of the particle-wave duality: 
real-space (appropriate for high-energy atomic-like excitations ) and momentum-space (appropriate for low-energy 
quasiparticle excitations) descriptions are reconciled within DMFT. 
    
Together with a representation of the local spectral function, 
the self-consistent effective atom (quantum impurity)  also allows for the calculation of a local self-energy. 
Within DMFT, this self-energy is identified with that of the full lattice or solid. This is precisely where an approximation is made, while 
in contrast the quantum impurity representation of the local spectral function can be viewed as exact.  
The frequency dependence of this local self-energy is in general quite complex, capturing both high-energy 
atomic multiplets and low-energy quasiparticles as well as a rich physics of crossovers between these entities as a 
function of energy and temperature. It is however, local in real-space, i.e 
momentum-independent\footnote{This statement applies to the self-energy in the local orbital basis. 
When upfolded to the Bloch basis in a real solid, the DMFT self-energy acquires momentum dependence.}.

This sounds like a very drastic approximation, and it is therefore a legitimate and important question to ask when this 
approximation is reasonable i.e when momentum dependence can be safely neglected. It is hardly possible to review this issue 
extensively here. Rather, this short note will present a few miscellaneous reflections. I will in particular 
emphasize that the local approximation becomes legitimate at high temperature or high energy. This is of course clear both 
on an intuitive basis and from the general principles of statistical mechanics, but this point of view has perhaps 
not been emphasized enough. Conceptually, it establishes DMFT as a sort of compass which can guide us towards the correct 
physical modelling on an intermediate energy scale . Of course, single-site DMFT must be supplemented by more sophisticated tools when 
dealing with lower energy scales/longer length scales.  
%

\section{Validating DMFT at high temperature: theoretical considerations and experiments with cold atoms.}

\subsection{Conceptual links between high-temperature expansions and mean-field theories}

Actually, any mean-field theory in statistical mechanics has strong conceptual links with the limit of high-temperature. 
In Ref.\cite{georges_yedidia_mft}, it was shown that mean-field theories of classical spin models can be viewed as a high-temperature 
expansion of the free-energy considered as a functional of the order parameter (obtained by a Legendre transformation from the 
energy in an external field), i.e an expansion in powers of 
$\beta J$ at fixed magnetizations $m_i$ (with $\beta=1/kT$ and $J$ the exchange coupling). 

To be specific, consider the Ising model $H=-\sum_{\langle ij\rangle} J_{ij} S_iS_j$. The free-energy functional $A$ can be written as:
\begin{equation}\label{eq:func_Ising}
-\beta A[\{m_i\};\beta J_{ij}] \,=\,\sum_i \Gamma_{\rm{loc}}[m_i] \,+\, \Gamma_{J} [\{m_i\};\beta J_{ij}]
\end{equation} 
In this expression, the local part of the functional corresponds to the entropy of independent spins with constrained magnetizations:
\begin{equation}\label{eq:entropy_Ising}
\Gamma_{\rm{loc}}[m_i]\,=\, -\left[ \frac{1+m_i}{2}\ln \frac{1+m_i}{2}+\frac{1-m_i}{2}\ln \frac{1-m_i}{2}\right]
\end{equation} 
and $\Gamma_J$ sums up all the non-local contributions to the functional due to the non-local interaction $J_{ij}$. 
Given the exact form of the functional $\Gamma_J$, the equilibrium magnetizations of the system are obtained from 
$\delta\Gamma/\delta\,m_i=0$ and hence are the solutions of $m_i\,=\,\tanh \frac{\delta\Gamma_J}{\delta m_i}$. 
From this equation, it is seen that $ \beta^{-1}\delta\Gamma_J/\delta m_i$ can be viewed as the {\it exact} self-consistent Weiss field, 
such that $m_i=\tanh \beta h_i$. 

The mean-field approximation for the Ising model with uniform couplings (e.g. ferromagnetic) $J_{ij}=J$ correspond to 
a truncation of the functional $\Gamma_J$ to its first-order contribution in $\beta J$, namely:
\begin{equation}\label{eq:MFT_Ising}
\Gamma_J^{\rm{MFT,ferro}}\,=\, \beta J\sum_{\langle ij\rangle}  m_i m_j
\end{equation}
Hence, the familiar mean-field equation $m_i=\tanh \beta\sum_j J_{ij} m_j$. This approximation becomes exact in 
the limit of large dimensions (lattice connectivity). In this limit, the exchange 
has to be scaled as $J=J^*/d$, and only the two terms (\ref{eq:entropy_Ising},\ref{eq:MFT_Ising}) survive in the large-$d$ limit. 

In contrast, there are situations in which additional terms of the high-temperature expansion of $\Gamma[\{m_i\}]$ must be retained in 
the large-$d$ limit. A most famous one is the case of a spin glass 
with a quenched distribution of exchange constants $J_{ij}$ of random signs. In this case, the second-order 
`Onsager reaction term' must be kept:
\begin{equation}\label{eq:MFT_SG}
\Gamma_J^{\rm{MFT,spin-glass}}\,=\, \beta \sum_{\langle ij\rangle} J_{ij} m_i m_j \,+\,
\frac{1}{2}\sum_{ij} (\beta J_{ij})^2 (1-m_i^2)(1-m_j^2)
\end{equation}
Minimization then yields the Thouless-Anderson-Palmer mean-field equations for the local magnetizations. 

Finally, there are situations in which an {\it infinite number of terms} in the high-temperature expansion must be summed up in order to 
obtain the correct free energy in the large-$d$ limit. One such example is the `fully frustrated' Ising model, a non-random model 
with ferromagnetic and antiferromagnetic bonds arranged in such a way that the product of bonds on each plaquette is 
negative. In this case, the exchange must be scaled as $J=J^*/\sqrt{d}$ in the large-$d$ limit. Among all possible high-temperature 
graphs, only a specific kind survive the large-$d$ limit (loops), but there are still an infinite number of them to be summed up because a 
loop of length $L$ contributes a factor $(\beta J^*)^L/d^{L/2}$ while there are of order $d^{L/2}$ such loops for large $L$. 
The resummation of this infinite set of graphs was performed by J.~Yedidia and the present author~\cite{yedidia_georges_fully_frustrated}, 
leading to a mean-field theory of the fully frustrated Ising model which is exact in the large-$d$ limit. 

Lattice fermion models share the same property than the fully frustrated Ising model, in that the fermion hopping 
must also be scaled as $t_{ij}=t^*/\sqrt{d}$ in the large-$d$ limit, as emphasized in the seminal article of Metzner and Vollhardt~\cite{metzner_vollhardt} 
This is in contrast to bosons in which the hopping would scale as $1/d$, in close analogy to a quantum XY model. 
Hence, an exact solution of interacting fermion models in large-$d$ corresponds to a resummation of an infinite 
number of terms in the high-temperature expansion, of a specific kind. 
Obviously, if I am allowed to bring in a personal recollection, this similarity with frustrated 
spin models played an important role in triggering my own interest (and that of G.Kotliar) in this problems, and our efforts to put it in 
the framework of a quantum generalization of mean-field theory~\cite{georges_kotliar_dmft}. 

The local observable is now the local Green's function $G_{ii}(\omega)$. In contrast to the local magnetization $m_i$, it is 
now a function of frequency (DMFT is a Green's function based theory of excited states, and energy dependence is crucial), and 
it should also be noted that $G_{ii}$ is not an order parameter, hence it is non-trivial at any temperature.  
In complete analogy with the above, a free-energy functional of $G_{ii}$ can be constructed fro the Hubbard model, 
separating local and non-local terms:
\begin{equation}
\Gamma[\{G_{ii}\}; t_{ij}]\,=\,\sum_i \Gamma_{\rm{imp}}[G_{ii}]\,+\,\Gamma_t[\{G_{ii}\}; t_{ij}]
\end{equation}
The first-term $\Gamma_{\rm{imp}}=\Gamma[\{G_{ii}\}; t_{ij}=0]$ is the free-energy functional of an Anderson impurity model, while $\Gamma_t$ is an (exact) 
kinetic-energy functional summing up all hopping processes. 
The expansion in the hopping $\beta t_{ij}$ is closely related 
to the high-temperature expansion~\cite{metzner_linkedcluster_prb_1991}. 
Minimization yields the {\it exact} hybridization function 
(quantum generalization of the Weiss field) $\Delta_{ii}(\omega)=\delta\Gamma_t/\delta G_{ii}(\omega)$. DMFT can be 
viewed as an approximation which retains only some hopping processes in $\Gamma_t$ 
(for an an explicit form of the DMFT approximation to $\Gamma_t$, see \cite{georges_strong}). 

\subsection{Benchmarking DMFT at high temperature}

\begin{figure}
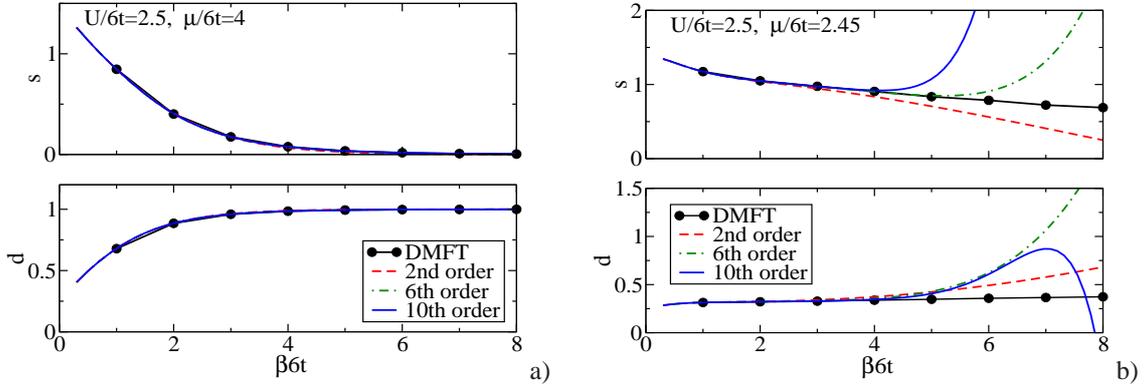

\includegraphics[width=68mm,height=50mm]{series_dmft_comp_lowdensity}~a)
\hfill
\includegraphics[width=68mm,height=50mm]{series_dmft_comp_intdensity}~b)
\caption{Comparison between DMFT and high-temperature series for the double occupancy 
$d=\langle n_{\uparrow}n_{\downarrow}\rangle$ and entropy per site as a function of inverse temperature $\beta 6t=6t/kT$, 
for $U/6t=2.5$. \textbf{a):} At $\mu/6t=4$ (corresponding to the high-density regime). 
\textbf{b):} At $\mu/6t=2.45$ (corresponding to an intermediate density $n\simeq 1.25$). 
Reproduced and adapted from Ref.~\cite{deleo_thermodynamics_pra_2011}}
\label{fig:series_dmft_comp}
\end{figure}

Recently, motivated by experiments on fermionic atoms in optical lattices, a systematic comparison between 
high-temperature series for the Hubbard model on a three-dimensional cubic lattice~\cite{Oitmaa2006} and the corresponding 
DMFT approximation has been performed~\cite{deleo_thermodynamics_pra_2011}. 
Two representative examples are reproduced on Fig.\ref{fig:series_dmft_comp}, which displays the 
double occupancy and entropy per site for an intermediate coupling (in units of 
the half-bandwidth) $U/6t=2.5$, as a function of inverse temperature. 
For low density of particles or holes (Fig.\ref{fig:series_dmft_comp}a) DMFT and the high temperature series are in excellent agreement with 
each other and allow to reach temperatures significantly smaller than the hopping amplitude. In this regime, 
it was actually shown~\cite{kozik_diagMC_epl_2010} that direct 
diagrammatic Monte Carlo simulations are in good agreement with DMFT down to $T/t=1/40$. 
At intermediate density ($n\simeq 1.25$ in Fig.\ref{fig:series_dmft_comp}b), it is seen that the series expansion breaks down already at a rather 
high temperature scale (significantly larger than the hopping amplitude). In the whole range of temperature where the series can be made sense of, agreement 
with DMFT is excellent. Direct benchmarking of the accuracy of DMFT below this scale is difficult in this regime of intermediate density and 
coupling used in Fig.\ref{fig:series_dmft_comp}b, because of the well-known limitations of direct lattice fermion simulations. 
Diagrammatic Monte Carlo studies at weaker coupling ($U/t\sim 4$)  in two dimensions~\cite{kozik_diagMC_epl_2010} 
have revealed deviations for $T \lesssim t$. 

In a nutshell, these comparisons fully support the intuitive view that physics become more local in space 
as temperature is raised (because the correlation lengths for any kind of short-range ordering become very short), so 
that single-site DMFT is a quantitatively accurate approach in those regimes. Because it involves a partial resummation of the high-temperature 
series, it can be used down to lower temperature than the series for a large range of densities, and furthermore allows for a direct calculation of  
frequency-dependent response functions. 

\subsection{Benchmarking DMFT in cold atom experiments}

A direct quantitative comparison with experiments in this high-temperature range is hardly possible in the 
solid-state context, and would anyhow be complicated by phonon degrees of freedom.
%
Experiments with cold atomic gases trapped in optical lattices offer however a beautiful opportunity.  
The rapid development of experimental techniques in this field allows for the engineering of 
artificial solids with a high degree of controllability, opening a new field of research to explore strongly correlated 
phases of quantum systems in a new context. 
Key steps along this path have of course been the theoretical proposal for realizing Hubbard-like models and 
observing the Mott transition in this context~\cite{jaksch_lattice_prl_1998}, and the experimental observation of the superfluid to Mott insulator 
transition of bosonic atoms~\cite{greiner_mott_nature_2002}. 
Recently, the crossover from an itinerant state into a Mott incompressible state as coupling is increased 
has been observed experimentally in the fermionic case as well~\cite{jordens_mott_nature_2008,schneider_mott_science_2008}. 
Those experiments have provided us with an `analog quantum simulator' validation of single-site DMFT, admittedly still 
in a rather high-temperature regime. 

\begin{figure}
\includegraphics[width=68mm,height=60mm]{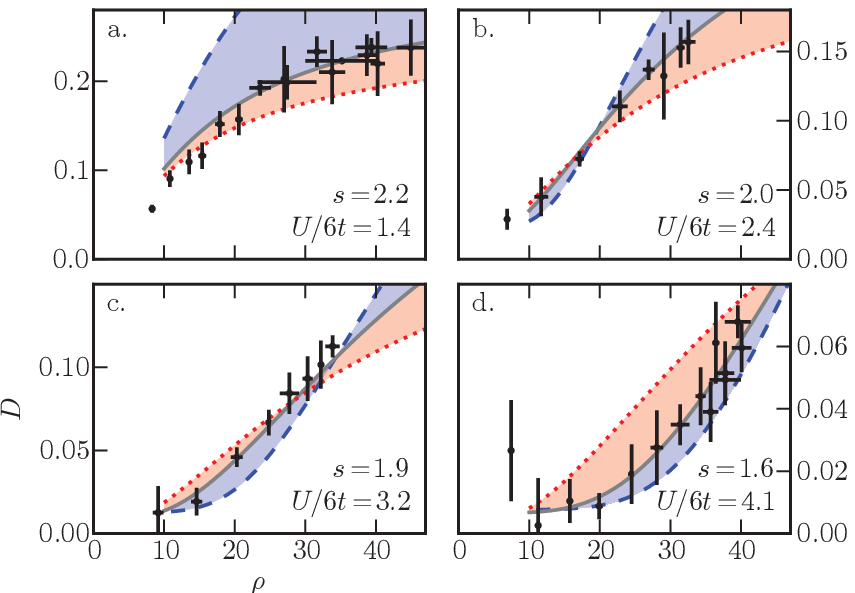}~a)
\hfill
\includegraphics[width=68mm,height=60mm]{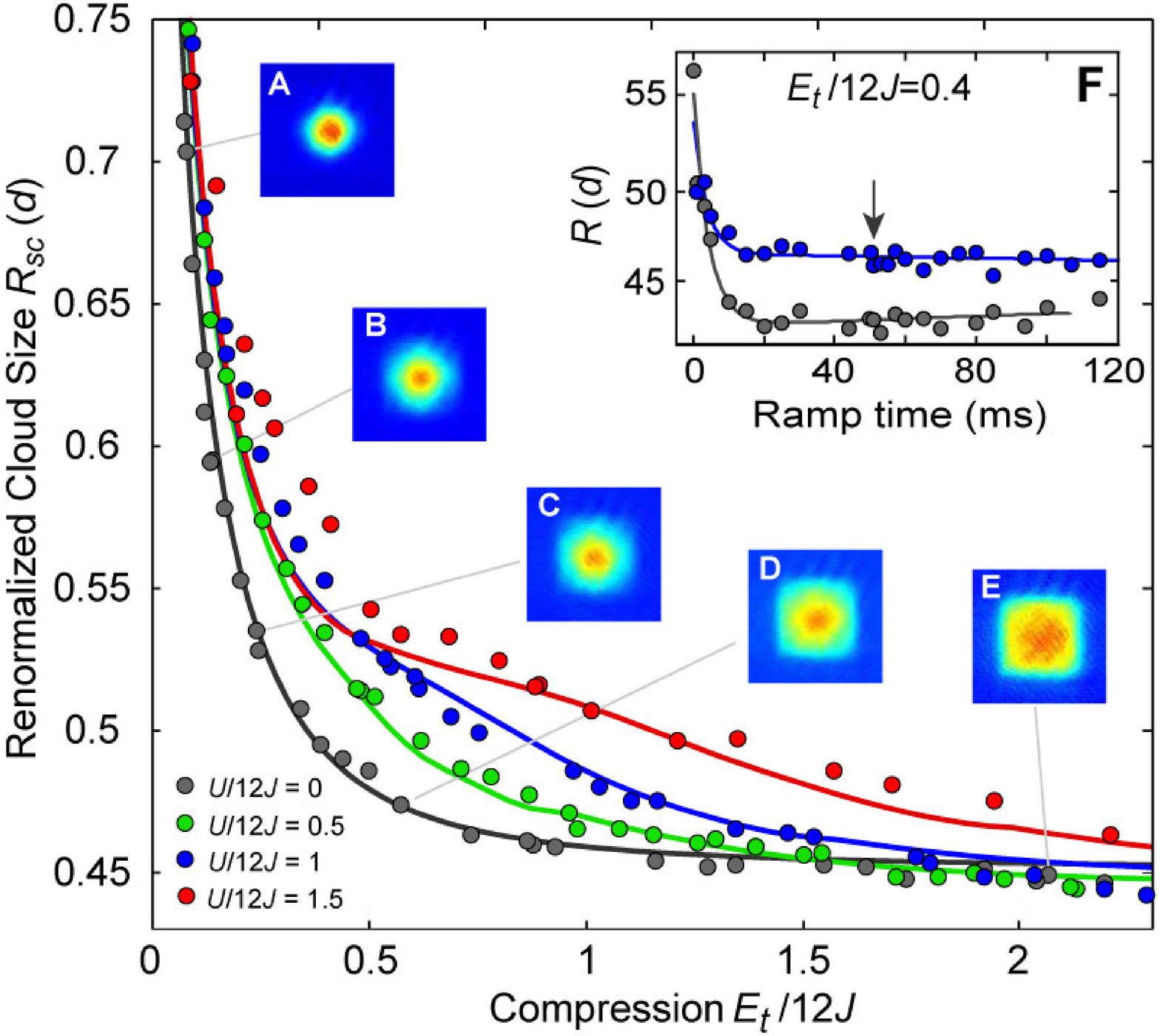}~b)
\caption{Experiments on cold fermionic atoms with repulsive interactions in a three-dimensional optical lattice. 
{\bf a)} (Reproduced from ~\cite{jordens_temperature_prl_2010}). 
    Double occupancy: experiment versus theory. Points and error
    bars are the mean and standard deviation of at least three experimental
    runs. The solid curve in each panel is the best fit of the second order
    high-temperature series to the experimental data and yields specific
    entropies of $s=2.2(2), 2.0(5), 1.9(4), 1.6(4)$ for the different
    interactions strengths of $U/6t=1.4(2), 2.4(4), 3.2(5), 4.1(7)$.
    Curves for $s=1.3$ (dashed curve) and $2.5$ (dotted curve) represent
    the interval of specific entropy measured before and after the ramping
    of the lattice. 
    {\bf b)} (Reproduced from Ref.~\cite{schneider_mott_science_2008}). 
    Cloud sizes versus compression. 
    Measured cloud size $R_{sc}$ in a $V_{lat} = 8\,E_r$ deep lattice as a function of the external trapping potential for various interactions $U/12t=0$ (black), $U/12t=0.5$ (green), $U/12t=1$ (blue), $U/12t=1.5$ (red) - in this figure the hopping is designated by $J$. Dots denote single experimental shots, lines the theoretical expectation from DMFT for $T/T_F=0.15$ prior to loading. The insets {\bf (A-E)} show the quasi-momentum distribution of the non-interacting clouds (averaged over several shots). {\bf (F)} Resulting cloud size for different lattice ramp times at $E_t/12t=0.4$ for a non-interacting and an interacting Fermi gas. The arrow marks the ramp time of 50\,ms used in the experiment.
    }   
\label{fig:coldatoms}
\end{figure}

Two such comparisons between theory and experimental results 
are reproduced in Fig\ref{fig:coldatoms}. In Fig\ref{fig:coldatoms}(a)~\cite{jordens_temperature_prl_2010}, 
the measured double occupancy as a function of atom number is 
compared to theoretical calculations performed at constant entropy (assuming that turning on the optical lattice 
corresponds to an adiabatic process). High-temperature series expansions was actually sufficient for this comparison, with DMFT 
yielding identical results. Fitting theory to experiment allows for a determination of the actual value 
of the entropy, and ultimately of the temperature attained after the lattice is turned on, for a given particle number. 
This analysis reveals that the lowest temperature that was reached in this experiment (at small atom number) 
is comparable to the hopping amplitude ($T\sim t$). The regime with very small double occupancy at the two largest values of $U/6t$ actually 
corresponds to the formation of a Mott plateau in the center of the trap. This is more clearly revealed in the 
measurement of the cloud size as a function of trap compression (Fig\ref{fig:coldatoms}b~\cite{schneider_mott_science_2008}) 
as a plateau signalling the onset of an incompressible regime for the largest value of $U$ displayed. 

Progress in cooling techniques will undoubtedly allow one to reach lower temperatures in future experiments. 
In this context,  it will be quite interesting to document deviations from single-site DMFT as longer-range 
correlations develop, and to benchmark theoretical calculations including these effects, such as cluster extensions 
of single-site DMFT (see below). 

\section{Spatial correlations}

As temperature is lowered, the range of spatial correlations (as measured by the correlation length for the various 
types of short-range order) increases, and corrections to single-site DMFT become increasingly important.  
Here, I briefly review some physical consequences, focusing on the physics associated with the proximity 
of a Mott transition.

\subsection{Entropy quenching}

Let me start with the description of the half-filled insulating state. At temperatures smaller than the Mott gap but much 
larger than the antiferromagnetic superexchange $J$  ($J\ll T \lesssim \Delta \sim U$), charge degrees of freedom (density fluctuations) 
are frozen, and we have a fluctuating paramagnet with a localized spin-$1/2$ on each site and 
very short-range spatial correlations. This is indeed the DMFT description of the paramagnetic insulator at strong coupling, and 
correspondingly a finite entropy per site is found in this regime $s\simeq \ln 2$. As temperature is lowered, a long-range 
ordered antiferromagnet will eventually be found below the N\'eel temperature $T_N$ (DMFT yields an upper bound for 
$T_N$ which, at strong cooupling, coincides with the classical mean-field estimate of the Heisenberg model). However, because 
short-range correlations are not included in the single-site DMFT description of the paramagnetic phase, the entropy remains 
frozen at $\ln 2$ throughout the paramagnetic phase, and only drops below $\ln 2$ for $T<T_N$. In reality, we expect short-range correlations 
of increasing range to develop well above $T_N$ (for $T$ of order a few times $J$) 
and the entropy to decrease gradually from the $\ln 2$ value as these correlations develop. 

\begin{figure}
\includegraphics[width=68mm,height=50mm]{entropy}~a)
\hfill
\includegraphics[width=68mm,height=50mm]{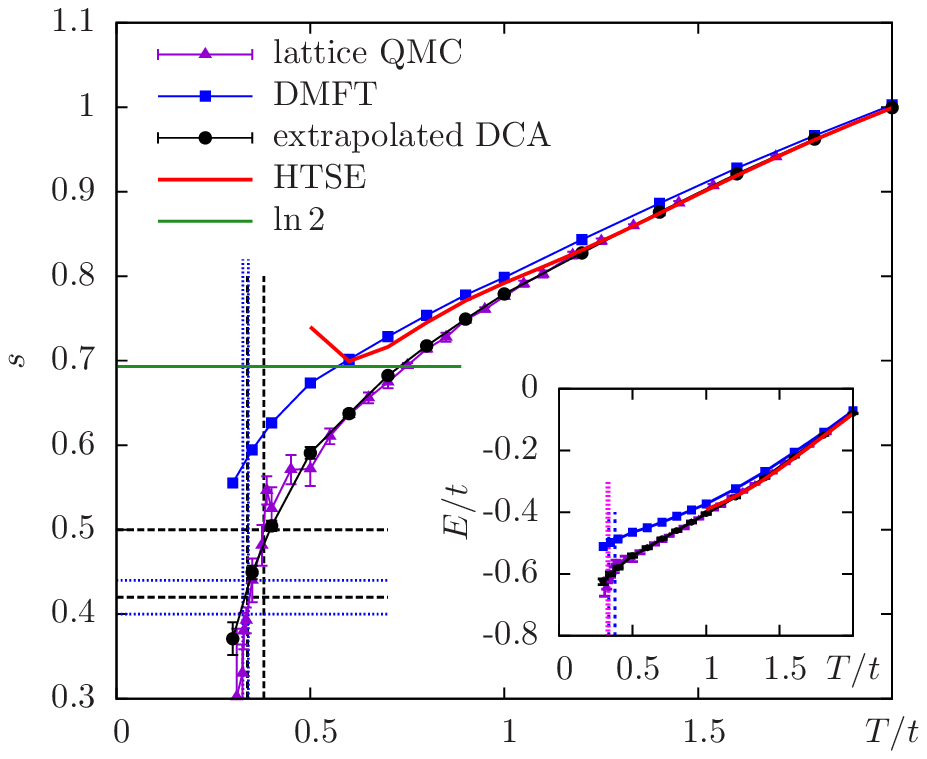}~b)
\caption{Entropy per site of the half-filled Hubbard model on a cubic lattice. 
{\bf a)} At $U/t=15$, from single-site DMFT and high-temperature series at various orders 
(reproduced from ~\cite{deleo_thermodynamics_pra_2011}). For comparison, the QMC result 
of \cite{wessel_entropy_prb_2010} for the Heisenberg model is also displayed. 
{\bf b)} At $U/t=8$, from single-site DMFT, extrapolations of cluster (DCA) methods, and direct 
lattice Monte-Carlo (reproduced from ~\cite{fuchs_thermodynamics_Neel_prl_2011}).
}
\label{fig:entropy}
\end{figure}

This is illustrated on Fig.\ref{fig:entropy}, which displays the entropy as a function of temperature for the half-filled 
Hubbard model on a three-dimensional cubic lattice. Fig.\ref{fig:entropy}a~\cite{deleo_thermodynamics_pra_2011} 
displays a stronger coupling $U=15 t$, for 
which the entropy saturation at $\ln 2$ within single-site DMFT is clearly visible. In this half-filled case, 
the high-temperature series are better behaved and can be trusted down to a rather low temperature ($T/6t \sim 1/15$).
The onset of short-range antiferromagnetic correlations clearly manifest themselves as a deviation of the entropy from the 
$\ln 2$ value. In Fig.\ref{fig:entropy}b~\cite{fuchs_thermodynamics_Neel_prl_2011}, 
a weaker coupling $U=8t$ is displayed which allowed for a detailed study using both direct 
lattice Quantum Monte Carlo calculations and a study using DCA cluster extensions of DMFT up to clusters of $64$ sites, which allowed 
for an extrapolation to the infinite-cluster size limit. At this weaker coupling, even the single-site DMFT entropy does decrease  below 
$\ln 2$ because we do not have a fully localized regime and density fuctuations are still important. Several lessons can be learnt from this 
figures, both on the range of temperatures where single-site DMFT can no longer be trusted due to the build-up of short-range 
correlations, but also on the ability of cluster extensions of DMFT to take into account these correlations accurately.  

The study of the entropy of the Hubbard model upon reaching the N\'eel transition has a direct relevance 
for the ability to eventually realize the antiferromagnetic state in cold atom experiments, as pointed out in~\cite{werner_cooling_2005}. 
There, it was emphasized that the relevant figure is the entropy per particle at the N\'eel transition, which was estimated 
from a Schwinger boson calculation to be of order $\ln 2/2$ for the half-filled Hubbard model on the cubic lattice, indeed quite 
close to the value obtained from the simulations in Fig.\ref{fig:entropy}b). Calculations extending these considerations to the 
system in an harmonic trap~\cite{deleo_trapping_mott_prl_2008,fuchs_thermodynamics_Neel_prl_2011} 
concluded that an entropy per particle of order $S/N\sim 0.65$ must be reached 
in order to reach the antiferromagnetic phase, corresponding to an intial temperature of the gas (assuming adiabatic switching of 
the lattice) of order $T_i/T_F =S/(N\pi^2) \sim 0.065$, roughly a factor of three colder than the temperatures that, to my knowledge, 
have been reported in the literature for those systems up to now (february, 2011).

\subsection{Pomeranchuk effect and the metal-insulator critical boundary}

\begin{figure}
\includegraphics[width=68mm,height=60mm]{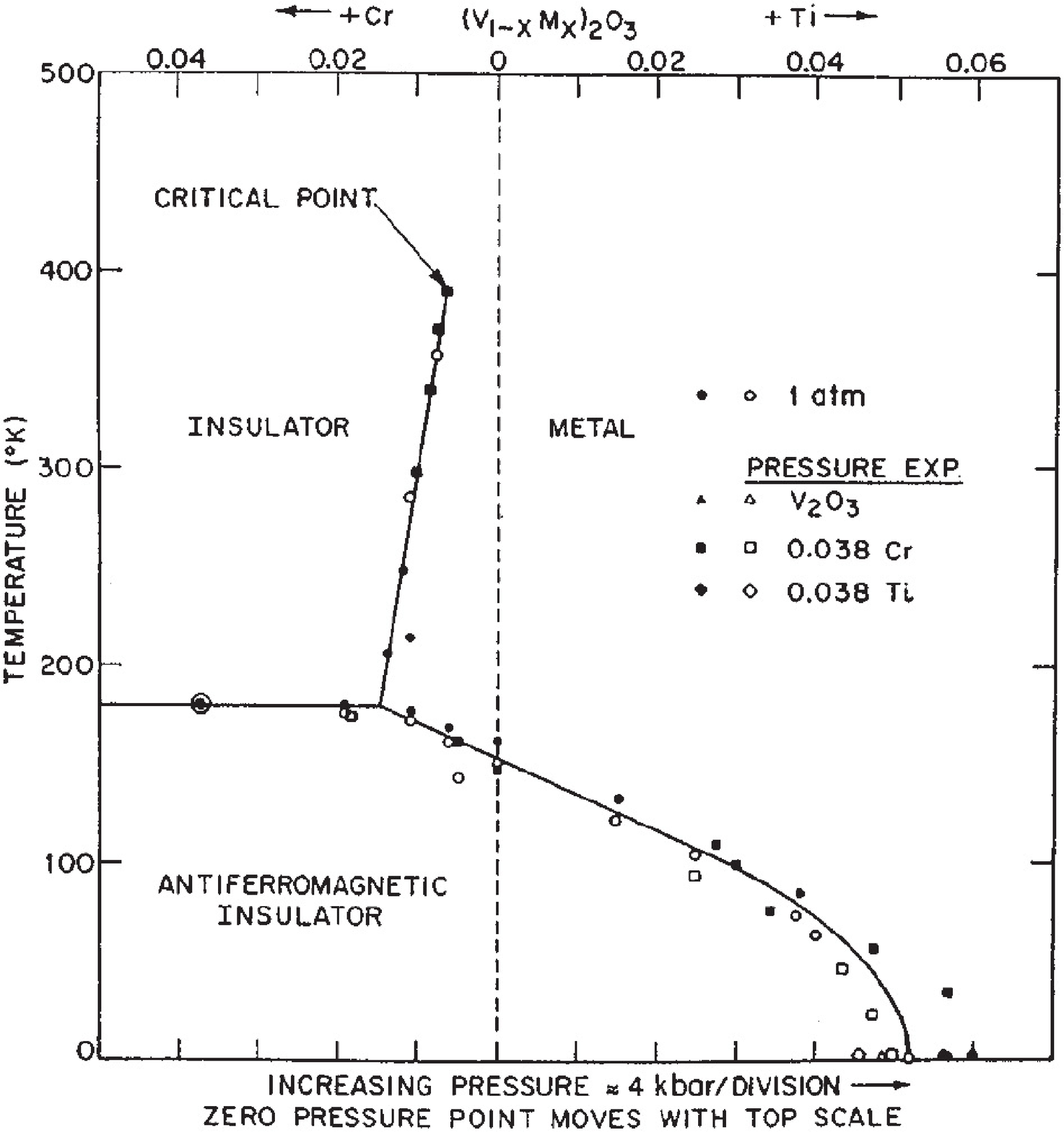}~a)
\hfill
\includegraphics[width=68mm,height=60mm]{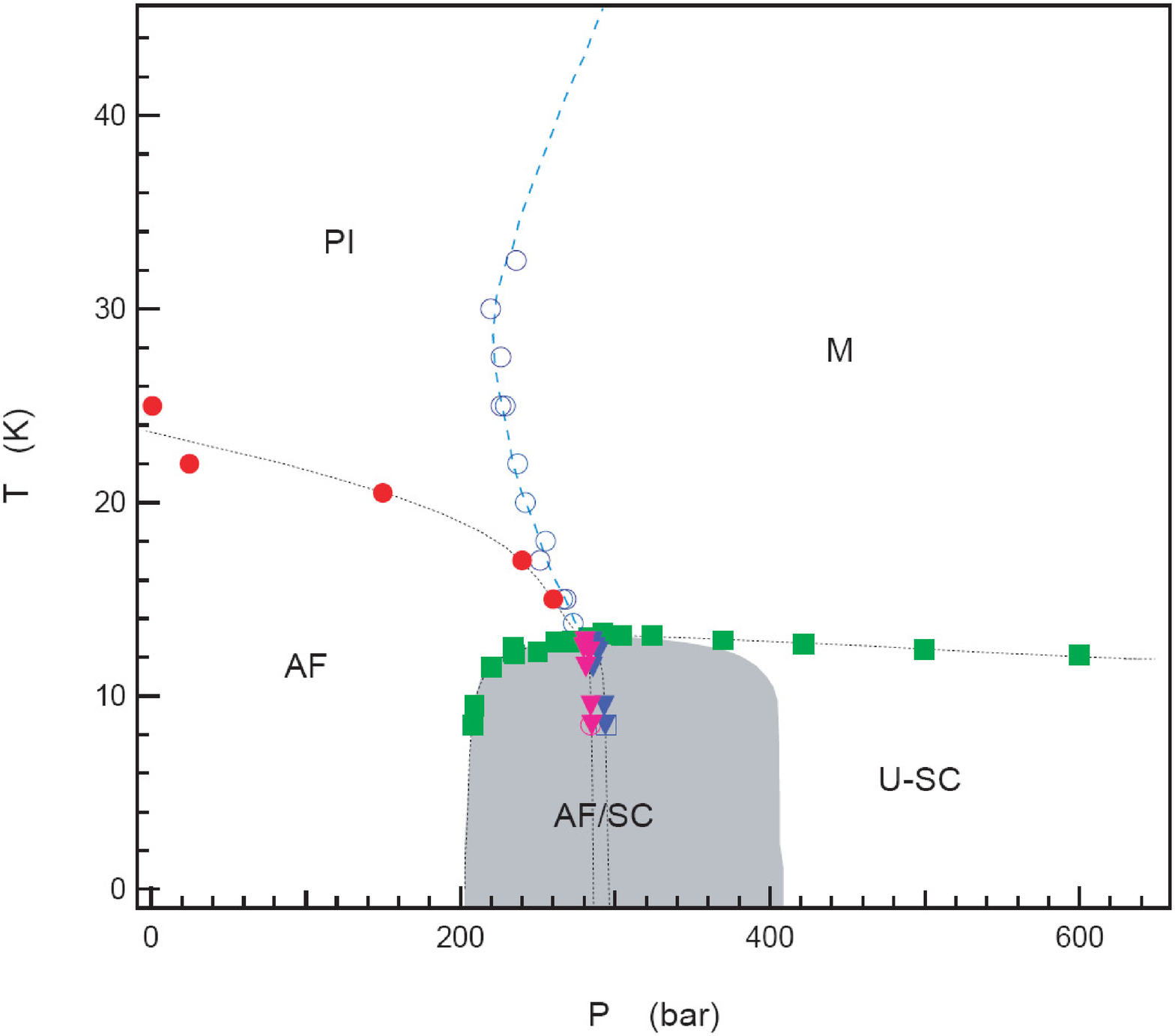}~b)
\caption{a): Phase diagram of (V$_{1-x}$ Cr$_{x}$)$_2$O$_3$ as a
function of either Cr-concentration $x$ or pressure (after\cite{McWhan73}).
Increasing $x$ by $1\%$
produces similar effects than {\it decreasing} pressure by $\sim 4$kbar, for this material.
b): Phase diagram of $\kappa$-(BEDT-TTF)$_{2}$Cu[N(CN)$_{2}$]Cl as a function of
pressure (after \cite{Lef00}).}
\label{fig:phasediag}
\end{figure}

The entropy of the insulating phase has also direct relevance for the shape of the critical boundary separating 
the metal from the insulating phase in materials which are poised close to the Mott transition. 
Indeed, the entropy in the metallic phase at low temperature is limited by the Fermi statistics of the quasiparticles and 
the Pauli principle, and hence linear in temperature $S \sim \gamma T$ (possibly with an enhanced value of 
$\gamma$, see below). As a result, if the insulating phase is a localized paramagnet with short-range 
spatial correlations and a high entropy per site, it becomes possible to gain a lot of free-energy by taking advantage of this entropy 
and localize the particles upon heating. This is the Pomeranchuk effect, which explains why liquid Helium 3 (a spin-$1/2$ 
fermion) can be solidified upon heating at high pressure, while this phenomenon is not observed for liquid Helium 4 
(a spinless boson). For this reason, the metallic phase of V$_2$O$_3$ can be turned into a Mott insulator upon heating, 
as manifested by the negative slope $dT_{\rm{MIT}}/dp >0$ of the (first-order) metal-insulator critical boundary as a function of 
pressure (Fig.\ref{fig:phasediag}a). 
This is in agreement with the observation of local moments in the insulating phase, with an enhanced Curie-Weiss 
law for the magnetic susceptibility. 
It is interesting to contrast this observation to the phase diagram of two-dimensional organics of the 
$\kappa$-BEDT family (Fig.\ref{fig:phasediag}b). These materials also display a first-order metal-insulator 
transition. In a restricted temperature range of temperature below the critical endpoint, the slope of the 
critical boundary is again positive, as for  V$_2$O$_3$, but turns around into $dT_{\rm{MIT}}/dp <0$ for $T\lesssim 30$K. 
This can be interpreted as due to antiferromagnetic correlations of increasing range, which quenches the entropy of the 
insulator so that the `Pomeranchuk effect' does not apply any more. 
This observation suggests that using single-site DMFT (while taking into account the actual electronic 
structure of the material) to describe the metal-insulator transition of  V$_2$O$_3$ at temperatures around the 
critical endpoint $T_c \simeq 450$K but also significantly below that scale is quite reasonable 
(as indeed has been done 
successfully~\cite{held_v2o3_prl_2001,limelette_v2o3_science,mo_V2O3_prominent_peak,keller_v2o3_prb_2004,poteryaev_v2o3_prb_2007}). 
In contrast, the range of applicability of single-site DMFT is probably more limited for the organic materials when considering the 
low pressure regime close to the insulating phase. 

\subsection{Orbital degeneracy and magnetic frustration favour local physics}  

It is interesting in this respect to note that V$_2$O$_3$ is a material in which all three $t_{2g}$ orbitals play an 
important role. Hence, there are orbital fluctuations which compete with magnetism and also induce frustration in 
the sign of the magnetic exchange couplings (hence the relatively low $T_N\simeq 170$K). In contrast, 
$\kappa$-BEDT are single-band materials (a remarkable rare situation among correlated materials, and also 
a unique feature of cuprates), although there is also geometric frustration due to the 
triangular crystal lattice.  

Quite generally, orbital degeneracy and magnetic frustration are  key factors in reducing the range of spatial 
correlations, and thus in extending the range of validity of the local approximation and of single-site DMFT. 
This is certainly one of the key reasons behind the successes that DMFT has had (in combination with electronic structure 
methods) with the description of the electronic structure of many materials including transition-metal oxides, 
rare-earth and actinide compounds (see e.g.~\cite{
kotliar_dmft_physicstoday,
kotliar_review_rmp_2006,
held_review_advphys_2007,georges_strong} for reviews). 

\subsection{Magnetic correlations and the effective mass of quasiparticles}

The quenching of entropy by magnetic correlations also has important consequences for the behavior of the 
effective mass of quasiparticles in the correlated metallic phase close to a Mott insulator. 

Within single-site DMFT, the quasiparticles in the metallic state  have a spectral weight $Z$ which vanishes as the 
insulator is approached (with $Z\propto \delta$ proportional to the doping level at large $U$, or 
$Z\propto U_c-U$ when approaching the insulator at commensurate filling). This in turn defines a coherence scale of the 
quasiparticles: $T^*\sim ZD$ (with $D$ a bare electronic energy scale, e.g. the half-bandwidth). Long-lived quasiparticles 
exist for $T<T*$, but become short-lived for $T>T^*$ and gradually loose meaning as a bad-metal regime is entered. 
At the coherence scale the entropy of the metal and that of the localized paramagnet become comparable: 
$\gamma T^* \simeq \ln 2$. This implies that $\gamma \propto 1/Z$ ($\propto \delta^{-1}, (U_c-U)^{-1}$), and hence that 
the quasiparticles have a heavy-mass close to the Mott transition: this is the Brinkman-Rice effect. Note that a purely local 
self-energy implies $m^*/m=1/Z$. 

When inter-site magnetic correlations are strong enough to quench out the entropy of the localized state, 
this argument is no longer valid however. In this case, we expect the divergence of the effective mass to be
cutoff. This is illustrated for example by the measured specific-heat coefficient of Sr-doped LaTiO$_3$, 
in which the Brinkman-Rice effect is indeed seen but cutoff by antiferromagnetic correlations at very low hole-doping.  
When the superexchange is large and efficiently binds neighboring spins into singlets, the mass 
enhancement need not be large at all. This is nicely captured by the simple formula $m^*/m \propto 1/(\delta+J/t)$ 
which can be derived within the large-$N$/slave-boson approaches to the $t-J$ model~\cite{kotliar_largeN_leshouches_1995}. 
It is seen that the Brinkman-Rice 
enhancement is cutoff for doping levels smaller than $\delta^*\sim J/t$, which precisely expresses that the superexchange is 
still active to quench the entropy as the quasiparticle coherence scale is reached ($J>t\delta$). 
In these approaches, the quasiparticle weight $Z\propto\delta$ becomes very different from $m/m^*\simeq J/t$ at low doping, 
which is only possible because the self-energy acquires momentum-dependence. 

These considerations teach us that the local approximation can in general not be sustained 
down to low doping levels, but should become more accurate as the doping is increased. Or more precisely, 
that at low doping the temperature scale below which it fails is rather high and set by the superexchange $J$. 
Below that scale, strong momentum dependence sets in.   
In contrast, at higher doping, the local approximation holds down to much lower temperatures, and momentum dependence 
is weaker (as indeed observed for overdoped cuprates). 

These considerations also imply that materials in which a large entropy is released as the quasiparticles are destroyed upon 
heating are likely to display rather short-range correlations and weak momentum-dependence, and hence are good candidates for a local 
DMFT description. Experimental signatures of this are a large quasiparticle effective mass but also large values of 
the Seebeck coefficient for $T>T^*$, which is directly related to the available entropy~\cite{chaikin_beni_thermopower_prb_1976}.  

\subsection{Singlet formation, pseudogap, nodal/antinodal dichotomy and cluster extensions of DMFT} 

In spin-$1/2$ one-band system with a large superexchange, the above effects are particularly significant, 
and a strong momentum dependence is expected at low doping $\delta\lesssim J/t$. This is indeed what is observed in the normal state 
of cuprates, in the temperature range $T_c<T<T^*$ where a pseudogap opens up in the antinodal region of the 
Brillouin zone, while reasonably well-defined quasiparticles survive in the nodal region. As doping is reduced 
towards the insulator, an increasingly large fraction of the Fermi surface is eaten up by the pseudogap: this is 
a quite different route to the Mott transition than the uniform reduction of quasiparticle weight following 
from local theories. 

\begin{figure}[t]
\includegraphics[width=\linewidth, height=6cm]{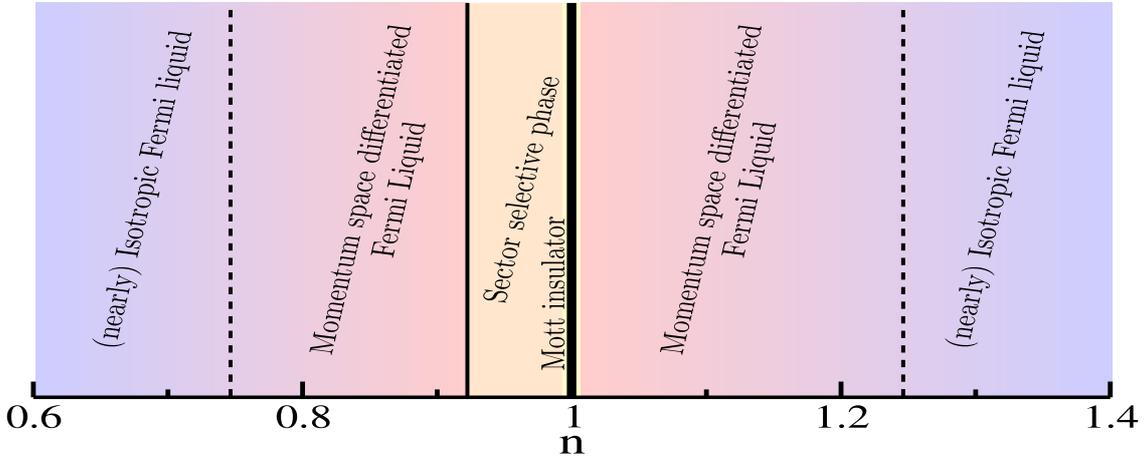}
\caption{The various doping regimes, from weak momentum dependence to 
strong momentum selectivity, as found in cluster extensions of DMFT (reproduced from \cite{gull_clustercompare_prb_2010}).}
\label{fig:regimes}
\end{figure}

Cluster extensions of DMFT have been quite successful at addressing this problem (for reviews of these approaches, 
and of the numerous works in the field over the past ten years, 
see e.g.~\cite{
maier_cluster_rmp_2005,
kotliar_review_rmp_2006,
tremblay_review_ltp_2006}). As for single-site DMFT, those approaches should be viewed 
as approaching the problem from the high-energy/high-temperature/high-doping side. Very low energies or small doping 
levels require very good momentum resolution which is hard to reach currently within those approaches. 
Nevertheless, robust qualitative trends can be established which do not depend strongly on the specific scheme or 
cluster size. In \cite{gull_clustercompare_prb_2010}, a comparative study of different clusters was performed in order to establish these robust 
qualitative trends which are summarized in Fig.\ref{fig:regimes}. At large doping levels, momentum dependence is weak and 
single-site DMFT is quite accurate. At intermediate doping, momentum differentiation emerges: on the hole-doped side, 
antinodal quasiparticles acquire a shorter lifetime than nodal ones. Finally, on the hole doped side, these 
approaches produce a transition at a critical doping below which the antinodal quasiparticles become gapped. This momentum-selective 
gapping is the cluster-DMFT description of the pseudogap. It is clearly associated with the formation of inter-site singlets, as 
can be checked by investigating the statistical weights of the different local states. Hence, at a qualitative level, 
those approaches seem to support some of the resonating valence-bond ideas, while extending them considerably by providing 
a theoretical framework in which the consequence of singlet formation can be studied in an energy and momentum dependent way. 

\section{Conclusion}

In this article, I have presented some reflections on DMFT and its extensions. The local approximation is a 
good starting point when spatial correlations are short-range, and this is favoured by any of the following: 
high-temperature, high-energy, high doping, large number of fluctuating degrees of freedom competing with each other, 
large orbital degeneracy, large degree of frustration. In those regimes, DMFT can be viewed as a compass providing 
guidance on the relevant degrees of freedom and their effective dynamics over intermediate energy scales. 
These intermediate energy scales and associated crossovers play a crucial role in the physics of strongly correlated materials. 

\begin{acknowledgement}
I am grateful to all the collaborators involved in the various projects referred to in this article, and especially to 
M.Ferrero, E.Gull, C.Kollath, G.Kotliar, A.J.Millis and O.Parcollet for inspiring discussions.  
Support from the Agence Nationale de la Recherche, the DARPA-OLE program and 
the Partner University Fund is acknowledged. 
\end{acknowledgement}

\def\bstname{adp}

%
%

\providecommand{\WileyBibTextsc}{}
\let\textsc\WileyBibTextsc
\providecommand{\othercit}{}
\providecommand{\jr}[1]{#1}
\providecommand{\etal}{~et~al.}

\end{document}